\documentstyle[draft]{mn}

\title[Testing the `clump' model of SiO maser emission]
      {Testing the `clump' model of SiO maser emission}

\author[M.\,D.\ Gray et al.]
       {M.\,D.\ Gray$^{1}$, R.\,J.\ Ivison$^{2}$,
        E.\,M.\,L.\ Humphreys$^{1,3}$ and J.\,A.\ Yates$^{4}$ \\ 
        $^{1}$ Department of Physics, University of Bristol, Tyndall Avenue,
               Bristol BS8 1TL\\
        $^{2}$ Institute for Astronomy, University of Edinburgh,
               Royal Observatory, Blackford Hill, Edinburgh EH9 3HJ\\
        $^{3}$ School of Chemistry, University of Bristol, Cantock's Close,
               Bristol BS8 1TS\\
        $^{4}$ Department of Physical Sciences, University of
               Hertfordshire, College Lane, Hatfield AL10 9AB }

\date{Accepted ... .
      Received ... ;
      in original form ...}

\pagerange{000--000}

\begin{document}

\maketitle

\begin{abstract}
Building on the detection of the $J=7-6$ SiO maser emission in both
the $v=1$ and $v=2$ vibrational states towards the symbiotic Mira,
R~Aquarii, we have used the James Clerk Maxwell Telescope to study the
changes in the SiO maser features from R~Aqr over a stellar
pulsational period. The observations, complemented by contemporaneous
data taken at 86\,GHz, represent a test of the popular
thermal-instability clump models of SiO masers. The `clump' model of
SiO maser emission considers the SiO masers to be discrete emitting
regions which differ from their surroundings in the values of one or
more physical variables (SiO abundance for example). We find that our
observational data are consistent with a clump model in which the
appearance of maser emission in the $J=7-6$ transitions coincides with
an outward-moving shock impinging on the inner edge of the maser zone.
\end{abstract}

\section{Introduction}

R~Aqr is a well-known symbiotic Mira, exceptional in that its envelope
supports maser emission (Martinez et al.\ 1988; Schwarz et al.\
1995). SiO maser emission has been detected in the low rotational
excitation transitions $J=2-1$ and $J=1-0$, and R~Aqr is also one of
only two symbiotic Miras to exhibit H$_{\rm 2}$O masers (22\,GHz:
Ivison, Seaquist \& Hall 1994, 1995; 321\,GHz: Ivison, Yates \& Hall
1998).

SiO maser emission in the $J=7-6$ rotational transitions of
vibrational states $v=1$ and $v=2$ had been predicted by theoretical
models (Doel et al.\ 1995; Gray et al. 1995) and, although
observations towards some of the brightest nearby masing objects
failed to detect these lines, they were detected towards R~Aqr (Gray
et al.\ 1996). Following this inital discovery, $J=7-6$ SiO masers
have been detected towards several stars of Mira, semi-regular
variable and supergiant types, proving that emission in these lines is
not based on the symbiotic nature of R~Aqr (Humphreys et al.\ 1997a).

The objective was to study the variation of the SiO maser
spectra, in both of the 300-GHz lines detected to date, over a stellar
pulsational period. This sequence of observations makes it possible to
investigate the validity of the `clump' model of SiO maser emission,
which considers the SiO masers to be discrete emitting regions which
differ from their surroundings in the values of one or more physical
variables (SiO abundance, for example). Positional considerations
dictate that data for some stellar phases will be unobtainable with
observations made during a single stellar period (over several stellar
periods, in fact, because the period of R~Aqr is close to 1\,yr) but
the nature of variability in SiO masers meant that it would be unwise
to augment the incomplete phase data with material drawn from the same
phase in subsequent periods.

Observational support for discrete SiO emitting regions has come from
recent VLBI studies of SiO masers at 43\,GHz (Diamond et al.\ 1994;
Miyoshi et al.\ 1994; Greenhill et al.\ 1995), where rings of maser
emission are resolved into individual maser features on the
milli-arcsecond scale. The spectra we observe are therefore a
superposition of the individual line shapes of many discrete objects,
in the same way that the molecular line profile of a galaxy is the
result of emission from many discrete molecular clouds. Very recent
VLBI observations of SiO masers (Boboltz et al.\ 1997) in the
circumstellar envelope of R~Aqr at 43\,GHz, using the VLBA, show that
the masers form a ring structure, similar to those found in other Mira
variables. Moreover, Boboltz et al.\ show convincingly that the ring
of 43-GHz masers around R~Aqr was contracting at an average speed of
4\,km\,s$^{\rm -1}$ between 1995 December 29 and 1996 April 05. By
coincidence, this time period overlaps with some of the observing
epochs used in the present work.

It is well known that SiO maser fluxes in the lower rotational
transitions vary considerably, both within the stellar period and
between different periods, in several SiO maser sources. R~Aqr is no
exception; variation in the $v=1$, $J=2-1$ transition is clearly shown
by Schwarz et al.\ (1995) and in the $v=1$, $J=2-1$ spectra in Fig.~1 of
the present work.

In the `clump' model, the variability of the SiO maser spectra is
interpreted as resulting primarily from the motion of the maser
regions as they follow the pulsations of the stellar envelope. As the
masing objects move, they experience changes in the radiation fields
incident upon them and may change in density and temperature. These
changes in physical conditions cause fluctuations in the maser
emission, which may be observed as a change in flux, or in the line
profile(s) which a particular object emits. An abrupt change in the
physical conditions will occur once per cycle due to passage through
an outward-moving shock front.

From our computational models (Humphreys et al.\ 1996; Humphreys et
al.\ 1997b), the number of spots which contribute significantly to a
SiO maser spectrum falls with increasing rotational excitation. On
this basis, we would expect clumps emitting in $J=7-6$ lines to be
much rarer than those emitting in, say, the $J=2-1$ lines. Our
prediction regarding the $J=7-6$ lines would therefore be of an almost
catastrophic form of variability, since the loss of two or three
dominant clumps could eradicate the entire spectrum in these
transitions and the emission could remain absent for some considerable
fraction of the stellar period. The loss of a similar number of clumps
would be likely to produce a more subtle change in the $J=1-0$ or
$J=2-1$ masers, which appear to be composed of several tens of
emitting clumps in most Miras (for example, Diamond et al.\
1994). R~Aqr appears to have a fairly typical number of 43-GHz SiO
maser spots (Boboltz et al.\ 1997).

A plausible mechanism for the formation of SiO maser clumps is through
thermal instabilites resulting from infrared band cooling by CO and
SiO (Cuntz \& Muchmore 1994). Thermal instabilites of this type result
in a bifurcated envelope structure with two phases which differ in
kinetic temperature (by typically $200-300$\,{\sc k}) and in molecular
abundance. One phase would form the maser clumps and the other a
hotter background medium.

The observations described here enable us to test this theory of the
origin of SiO maser clumps by measuring the lifetimes of maser
features in the observed spectra against the timescales predicted by
Cuntz \& Muchmore (1994). The $J=7-6$ lines are particularly valuable
for this task, since they are likely to arise from the contributions
of only a few emitting clumps (as discussed earlier).  New spectral
features should appear on the cooling timescale ($<10^6$\,s) and
should last until a clump is re-heated by the passage of a
shock-wave. Shocks would be likely to modify each clump approximately
once per pulsational period, or on a timescale of $\sim 3.3 \times
10^7$\,s, in the particular case of R~Aqr, which is much larger than
the cooling time.

\begin{figure*}
\begin{center}
\setlength{\unitlength}{1cm}
\begin{picture}(17.5,22.35)
\put(0,0){\includegraphics{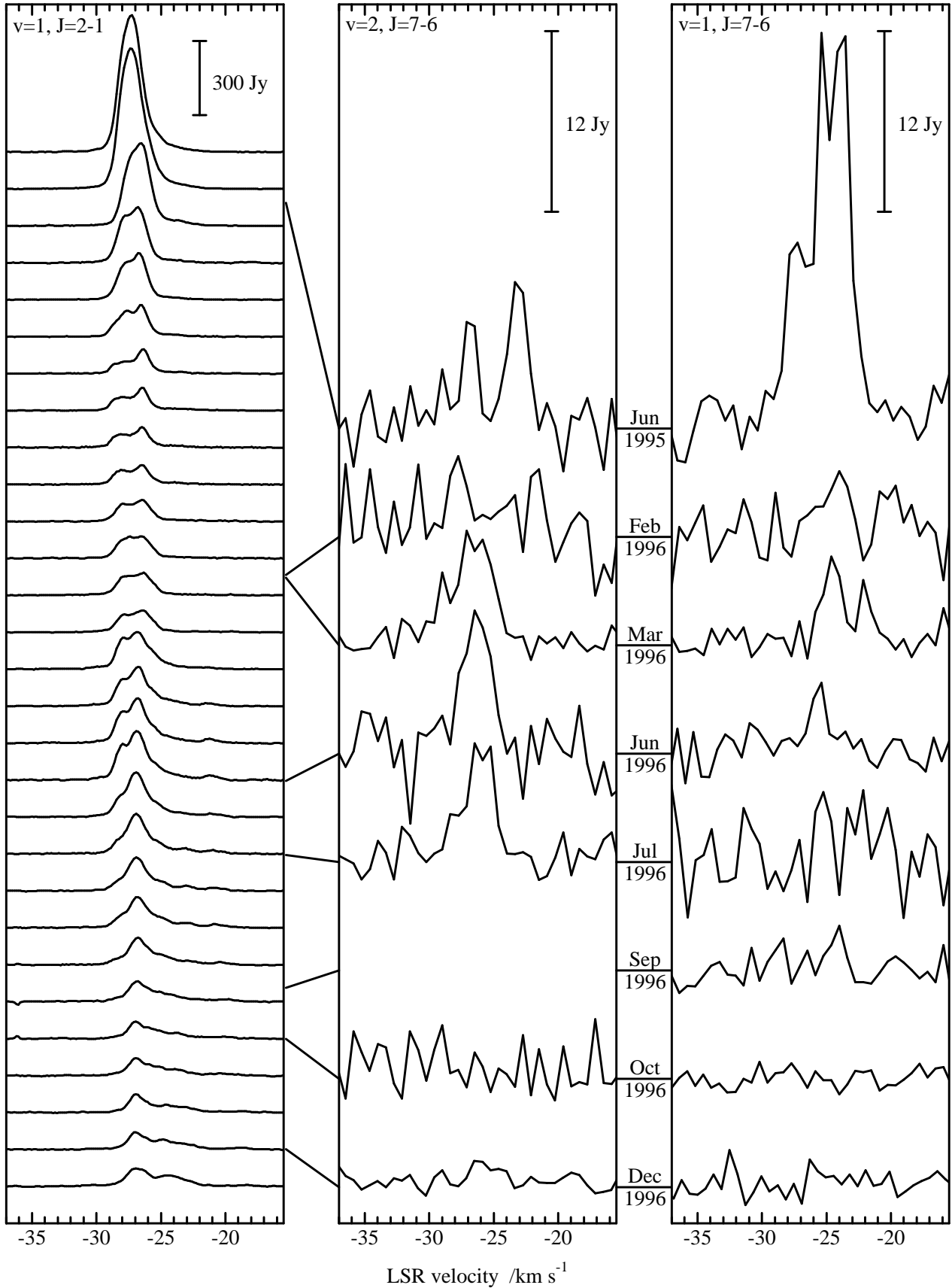}}
\end{picture}
\end{center}
\caption{R~Aqr SiO maser transitions obtained during the course of
1995---96 (top to bottom). Left: $v=1,J=2-1$ maser (86\,GHz); centre:
$v=2, J=7-6$ maser (299\,GHz); right: $v=1,J=7-6$ maser
(301\,GHz). Lines between the panels indicate data that are
contemporaneous.}
\end{figure*}

\section{Observations}

The 15-m James Clerk Maxwell Telescope (JCMT), Mauna Kea, Hawaii, was
used over a period of one year to monitor the $J=7-6$ SiO lines ($v=1$
and $v=2$) towards R~Aqr. The rest frequencies used were 301.814\,GHz
for the $v=1$ transition and 299.704\,GHz for the $v=2$ transition,
with a velocity of $-25.0$\,km\,s$^{-1}$ relative to the LSR. The
liquid-helium-cooled, single-channel SIS mixer receiver B3i was used,
with a digital autocorrelation spectrometer backend providing a
bandwidth of 250\,MHz with channel spacings of 0.125\,MHz (later
binned to 0.625\,MHz or 0.63\,km\,s$^{-1}$).  The $v=1$ and $v=2$
lines were tuned in the upper and lower sidebands, respectively. The
time spent on source (inclusive of overheads for sky subtraction,
exclusive of other overheads, such as the time spent nodding the
telescope to alternate the signal and reference beams) varied between
1200 and 5400\,s. Values of $T_{\rm rec}$ were around 200\,{\sc k},
whilst $T_{\rm sys}$ varied between 500 and 1400\,{\sc k}. JCMT data are
summarised in Table~1.

\begin{table}
\caption{High-frequency SiO maser data from the JCMT.}
\begin{tabular}{lrr}
UT Date       & Line Frequency & Peak Flux \\
              & /GHz           & /Jy       \\
1995 Jun 08   & 299 & 9.7     \\
              & 301 & 26      \\
1996 Feb 27   & 299 & 5.4     \\
              & 301 & 4.4     \\
1996 Mar 15   & 299 & 7.6     \\
              & 301 & 5.9     \\
1996 Jun 28   & 299 & 9.5     \\
              & 301 & 4.7     \\
1996 Jul 27$^{\dagger}$  & 299 & 7.9     \\
              & 301 & 4.8     \\
1996 Sep 15   & 301 & $<$3.0  \\
1996 Oct 06   & 301 & $<$1.0  \\
1996 Oct 22   & 299 & $<$3.6  \\
1996 Dec 01   & 299 & 1.7     \\
              & 301 & $<$1.9  \\
\end{tabular}

\vspace*{1mm}
\noindent
$^{\dagger}$ indicates an average, at 299\,GHz, of two close
epochs (1996 July 27 and 29). 
\end{table}

\begin{table}
\caption{SiO maser data from SEST.}
\begin{tabular}{lrr}
UT Date       & Peak Flux & Integrated Flux   \\
              & /Jy       & /Jy\,km\,s$^{-1}$  \\
&&\\
1995 May 25   & 553  & 1300    \\
1995 Jun 02   & 570  & 1400    \\
1995 Jul 07   & 335  &  830    \\
1995 Aug 13   & 226  &  580    \\
1995 Aug 25   & 189  &  470    \\
1995 Sep 21   & 128  &  340    \\
1995 Oct 24   & 98.9 &  240    \\
1995 Oct 30   & 92.6 &  240    \\
1995 Nov 07   & 82.6 &  220    \\
1995 Nov 15   & 78.9 &  230    \\
1995 Dec 02   & 86.9 &  260    \\
1995 Dec 15   & 89.1 &  290    \\
1996 Jan 04   & 91.9 &  290    \\
1996 Jan 18   & 88.0 &  270    \\
1996 May 01   & 151  &  480    \\
1996 May 23   & 159  &  470    \\
1996 Jun 10   & 181  &  540    \\
1996 Jun 28   & 198  &  590    \\
1996 Jul 13   & 181  &  510    \\
1996 Jul 30   & 163  &  470    \\
1996 Aug 07   & 136  &  380    \\
1996 Aug 21   & 124  &  380    \\
1996 Sep 06   & 109  &  320    \\
1996 Sep 21   & 82.1 &  300    \\
1996 Oct 07   & 69.8 &  220    \\
1996 Oct 16   & 71.1 &  220    \\
1996 Nov 22   & 74.6 &  230    \\
1996 Dec 05   & 69.2 &  250    \\
1996 Dec 27   & 73.3 &  300    \\
\end{tabular}
\end{table}

\addtocontents{lot}{SEST}

R~Aqr, supporting a very bright SiO maser at 86.243\,GHz, is regularly
used as a pointing source at the 15-m Swedish-ESO Submillimetere
Telescope (SEST), La Silla, Chile. Data from short integrations
(300\,s, typically) taken during the period between 1995 June and 1996
December using the dual-polarization Schottky receiver and the
acousto-optical high-resolution spectrometer (43-kHz channel
spacings) were kindly made available to us (L.-\AA.\ Nyman, private
communication). SEST data are summarised in Table~2.

\section{Results and Discussion}

\begin{figure*}
\begin{center}
\setlength{\unitlength}{1cm}
\begin{picture}(17.5,14.3)
\put(0,0){\includegraphics{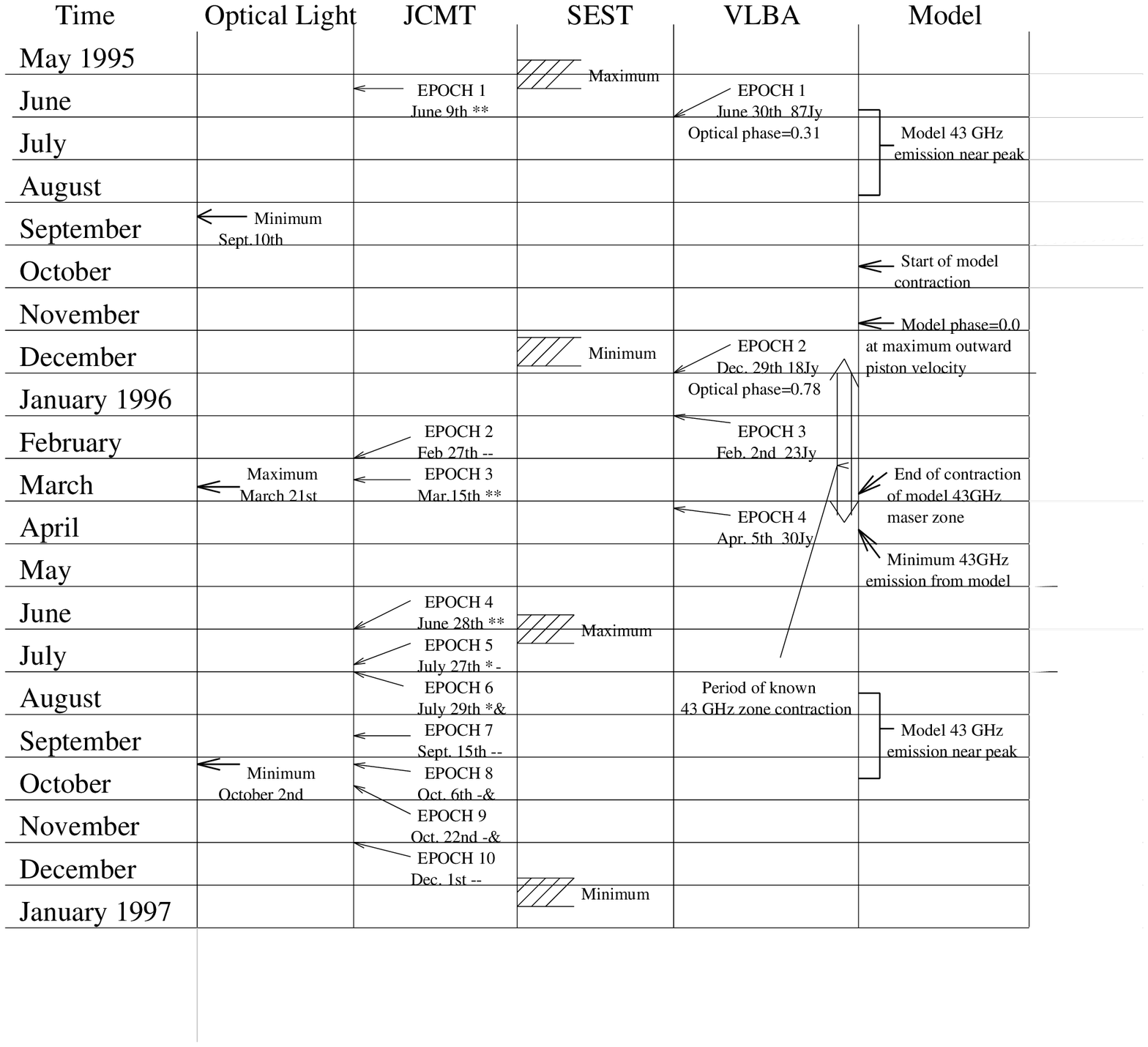}}
\end{picture}
\end{center}
\caption{Timeline for the observations discussed in this work together
with model data used in their interpretation. Symbols following the
epoch date in the JCMT column have the following meanings: *
observed and detected; -- observed and not detected; \& no
observation. The first symbol relates to 299\,GHz and the second to
301\,GHz in each case.}
\end{figure*}

Fig.~1 shows the observed spectra of the SiO $J=2-1$ and $J=7-6$ lines
towards R~Aqr. These data were reduced in the standard manner, using
{\sc specx} V6.7 (Padman 1993). Data reduction included re-binning to
produce channels of width 0.63\,km\,s$^{-1}$ (from five observational
bins in the case of the $J=7-6$ lines). Conversion from the $T_{\rm
A}^*$ scale to flux-density units was achieved using a conversion
factor of 30\,Jy\,{\sc k}$^{-1}$ for the data from the JCMT and
25\,Jy\,{\sc k}$^{-1}$ for data from SEST.

Inspection of Fig.~1 shows that all the SiO maser lines observed are
highly variable over time, as expected.  A comparison of the $J=2-1$
and $J=7-6$ spectra shows that the $J=7-6$ spectra are probably more
variable than the $J=2-1$ spectra. However the comparatively low
number of epochs available for the high-frequency observations reduces
the significance of this conclusion. It is also clear that whilst the
$J=2-1$ transition shows considerable variation in intensity over the
duration of the experiment (a factor of 8.33 in peak intensity between
the brightest and weakest epochs), the $J=2-1$ maser never disappears
entirely. On the contrary, the $v=1, J=7-6$ line appears to be
entirely absent, to a $3$-$\sigma$ level of $<1$\,Jy during 1996 October,
a factor of at least 25 lower than the largest peak in this
transition. The $v=2, J=7-6$ line is probably also absent for part of
the experimental period, but higher noise makes this less
statistically significant.

The shape of the high-frequency spectra for both lines is highly
variable.  It is clear from Fig.~1 that the line profiles have changed
significantly between 1996 March 15 and June 28 and again between the
latter date and July 27. Only over the shortest timescale, two days
between 1996 July 27 and 29, was the 299-GHz line profile very similar
and these two spectra have been averaged in Fig.~1. There was no
301-GHz observation on July 29. This variability suggests that the
clumps supporting high-frequency masers have lifetimes of between 2
and 30\,d.

R~Aqr has a pulsational period of 387\,d. From the AAVSO phase data in
Boboltz et al.\ (1997), the period covered by our data contains two
optical minima and one maximum. From these same phase data, we can fix
the optical maximum to 1996 March 21 to an accuracy of 1\,d. We note
that the light curve of R~Aqr is slightly skewed, with the rise from
minimum to maximum taking only 0.42 of a cycle (Kholopov, 1985). This
information allows us to fix the minima, which are less well defined
than maxima in R~Aqr, to 1995 October 11 and 1996 November 02. The minima,
however, are not used as timing markers for synchronization with other
data and are much less important than the maxima in the present work.
We also know that the envelope region containing the 43-GHz masers was
shrinking between late 1995 December and early 1996 April. A likely sequence
of events can be constructed from a combination of our data, the work
of Boboltz et al.\ and the hydrodynamic models of Bowen (1988, 1989)
augmented by maser models as in Humphreys et al.\ (1996). The
behaviour of these various aspects of the envelope are summarized in
the composite timeline shown in Fig.~2.

Both the 300-GHz lines were probably present on 1996 February 27, but
detections were marginal because of high noise levels. Both masers
were detected on 1996 March 15. These same masers were then detected on
three more occasions in 1996 June and July (a 301-GHz observation was
not attempted on the last of these occasions). These were the final
strong detections: further observations in 1996 September, October (2 sets) and
December failed to detect either line, except for a marginal detection
(1.7\,Jy) of the 299-GHz line in 1996 December. It is unfortunate that
additional data could not be taken in 1996 April or May, but it appears
to be the case that 300-GHz masers were active much of the time
between 1996 March and the end of 1996 July. These were then destroyed, by
passage through a shock or by some other adjustment to prevailing
physical conditions unsuitable for $J=7-6$ emission, at some time in
1996 August or early September and never re-formed during the remainder of the
experiment.

If we now consider the hydrodynamic situation, the zero model phase
(maximum expansion velocity of the sub-surface `piston' into the
photosphere) occurs at visual phase $\sim 0.7$ (Humphreys et al.\
1996), which in this case would be either in late 1995 November, or
just after the final set of JCMT observations in mid-1996
December. The period for which the model 43-GHz maser zone is in
contraction lasts from model phase $\sim0.85-0.90$ to model phase
$\sim0.25-0.35$. Masing clumps with larger than average radial
positions tend to contract, in modelling results, until the larger
phase value and vice-versa (Humphreys et al.\ 1997b). It can be seen
from Fig.~2 that the model and observed contraction zones are
compatible to within $\sim0.05$ of a complete cycle (17\,d) provided
that the observations performed by Boboltz et al.\ represent the
latter stages of the 43-GHz zone contraction. If this synchronization
of the model and the VLBA data is adopted, a consequence is that a
small phase-lag of $0.1-0.2$ of a period should appear between the
SEST (86-GHz) data and the VLBA (43-GHz) data. Unfortunately there are
only four VLBA epochs, so this prediction cannot be tested exactly,
but the 87-Jy VLBA signal shortly after the first SEST maximum in
Fig.~2, compared to the much lower fluxes measured during the
contraction period (18---30\,Jy) indicate that the data are not
inconsistent with a small phase difference between the 43- and 86-GHz
cycles, with the 43-GHz cycle leading. The position of the cycle
minimum in the model output is considerably less well defined than the
maximum and could occur somewhat earlier, but not later than its
marked position in Fig.~2.

Turning now to the 300-GHz data, there appears to be a significant
phase-lag between the $J=7-6$ maser emission and both the
low-frequency lines, assuming a 300-GHz peak somewhere between mid-March
1996 and late Jun 1996 (model phase between $\sim0.35$ and
$\sim0.65$). In fact, the lag with respect to the 43-GHz emission
would amount to almost half a cycle. The large phase lag of the
300-GHz masers with respect to those at low frequency means that the
data are not consistent with the idea that objects masing at 300\,GHz
are destroyed by the passage of a shock wave through the maser zone:
instead, they must be rapidly created in its aftermath and die away
for an entirely different reason. Passage of the shock through the
43-GHz maser zone can be timed observationally, to early April at the
earliest, since it would end the episode of infall.  However, data
from Boboltz et al.\ represent a broad average over many maser
features and the model data show that individual spots should be
struck by the shock as early as late February or as late as late April,
depending on the mean radial position of the spot: spots with smaller
mean radial positions suffering impact by the shock at earlier
times. The 300-GHz masers become strong between 1996 February 27 and March
15. If the clumps which give rise to this high-frequency emission are
those which lie, on average, closest to the star, then they could be
produced by shock impact in late February or early March, followed by a
period of cooling to below 4500\,{\sc k} when thermal instabilities
allow the formation of clumps. It is noted that model predictions are
that the vast majority of $J=7-6$ emission is produced by clumps with
temperatures between 3000 and 5000\,{\sc k}, consistent with clumps
formed by the action of the higher density regions of Cuntz \&
Muchmore's `CO Instability Island' for a stellar temperature of
3000\,{\sc k}.

\subsection{Timescales}

Three timescales are important in determining the behaviour of the
post-shock material: the hydrodynamic time, $\tau_{\rm H}$, the
radiative cooling time, $\tau_{\rm R}$ and the chemical time
$\tau_{\rm C}$. The hydrodynamic time is the timescale on which the
post-shock gas will cool if there is negligible radiative cooling. In
the inner regions of a Mira envelope, where overall expansion is a
small perturbation on the shock-wave and gravity driven cyclic
motions, the formula for $\tau_{\rm H}$ in Cuntz \& Muchmore (1994)
reduces to $\tau_{\rm H}$=${\rm c_{m}P}$/${\rm 2u_{2}}$ where P is the
stellar pulsational period, c$_{\rm m}$ is the average sound speed in
the envelope at a given radius and u$_{\rm 2}$ is the velocity of the
post-shock gas. The speed, c$_{\rm m}$, is a weak function of radius
and a mean value of 3.4\,km\,s$^{\rm -1}$ has been used for the whole
maser zone.

The radiative cooling time is the timescale on which the post-shock
gas will cool back to the local radiative equilibrium temperature,
provided that the most efficient coolant species, for a given
temperature and composition, are present in the gas.  The presence of
the coolant species depends on the chemical time. Cooling times given
in Cuntz \& Muchmore (1994), based on CO cooling, with a small
contribution from SiO, agree well with those predicted by the more
sophisticated cooling function developed in Woitke et al.\ (1996). The
good agreement relies on the fact that the total cooling time is
dominated by the slowest phase, between about 7000\,{\sc k} and the
instability temperatures of 4000---4500\,{\sc k} which is also dominated
by CO infrared band cooling.  We note that if efficient radiative
cooling applies, the shocks will be closer to isothermality than those
in the Bowen models, with the result that the enhanced temperature
regions behind the shocks would be markedly thinner than those which
appear in, for example, Fig.~1a of Humphreys et al.\ (1996).

The chemical time is the most difficult of the three timescales to
evaluate because it depends on a complex reaction network: it is the
longest of a set of timescales, each of which is a typical time taken
to re-establish the equilibrium abundance of an important
species. In the present work, the important species are CO (for
cooling), SiO (for masing and cooling) and H$_{\rm 2}$ (because the
collisional rate coefficients used in the maser modelling are a better
representation of collisions between SiO and H$_{\rm 2}$ than between
SiO and H). Chemical timescales are not calculated in any detail in
Cuntz \& Muchmore and we have adopted the more accurate values
calculated in Beck et al.\ (1992). We have used the standard Model~A
from Beck et al.\ with the modified temperature structure from Wirsich
(1988). High-temperature post-shock regions mean that there is
sufficient radiation to drive some photochemistry, making Model~A the
better approximation to our system for most parameters, whilst the
temperature structure from Wirsich is a better approximation to the
Bowen data than the monotonic decay from 1100\,{\sc k} at 2\,R$_{*}$
used in the standard Model~A. The timescale for CO is the longest at
all radii of interest, so this is taken as the chemical time. All
three timescales are tabulated for post-shock conditions in Table~3.

\begin{table}
\caption{Conditions in the post-shock gas as a function of model phase,
$\phi$.}
\begin{tabular}{lcccccc}
$\phi$ & R$_{\rm S}$ & $\tau_{\rm H}$ & $\tau_{\rm R}$ & $\tau_{\rm C}$ &
n$_2$   & n$_2$/n$_1$\\
      &  /R$_{*}$   & /d         &  /d        & /d         &
/cm$^{-3}$ &  \\
\\
0.0  & 1.25 &   67 & 0.17  & 0.005  & 7.9(12) & 140 \\
0.25 & 1.83 &   80 & 0.59  & 0.012  & 7.1(10) &  28 \\
0.5  & 2.40 &   99 & 0.97  & 0.058  & 1.0(10) &  13 \\
0.75 & 2.80 &   95 & 1.50  & 0.183  & 2.5(9)  &  11 \\
\end{tabular}

\vspace*{1mm}
\noindent
Note: The bracket notation has the mantissa outside the bracket and
the decimal exponent inside so, for example, $7.9(12) = 7.9 \times
10^{12}$.  R$_{\rm S}$ is the radial position of the shock front,
n$_1$ is the pre-shock density and n$_2$ the post-shock density after
the gas has cooled to 4500\,{\sc k}. Initial post-shock temperatures
lie between 8000 and 10000\,{\sc k} for all phases in the table.
\end{table}

It can be seen from Table~3 that the chemical timescale is the
shortest of the three timescales at all radii, so we expect that the
necessary coolant and masing molecules will be regenerated following
the passage of a shock before significant cooling occurs by
expansion. Cooling should therefore occur on the radiative timescale
because it is, firstly, markedly shorter than the timescale for
hydrodynamic cooling, and secondly, considerably longer than the
chemical time required to reform the coolant molecules dissociated by
the passage of the shock. This latter condition renders it unnecessary
to include a contribution for the chemistry in the cooling time.  A
significant delay appears between the model phase zero and the
appearance of the first 300-GHz masers. This is almost certainly
because the density of the post-shock material (see Table~3) is too
high to allow the necessary population inversions to develop before a
phase of about 0.25. Once the post-shock conditions are suitable,
maser clumps can develop via cooling instabilities on a timescale of
1---2\,d.  The 300-GHz masers would be the youngest masing objects,
forming at the highest instability temperatures of around
4000---4500\,{\sc k} (Cuntz \& Muchmore 1994). The lifetimes of
300-GHz maser clumps would be governed by a period of slower cooling
towards the inter-shock temperature of the envelope ($\sim$1500\,{\sc
k}) on a timescale of 5---20\,d, consistent with the variability of
the high-frequency spectra (Fig.~1).

Further evolution of the circumstellar shell probably follows the
sequence described here: the original 300-GHz emitting clumps cool
further and begin emitting instead in lower frequency lines, whilst as
the shock progresses outwards through the envelope it destroys
existing low-frequency emitting clumps and generates new
high-frequency emitters. Eventually, beyond a model phase of about
$0.70-0.75$, the shock becomes too weak to generate new clumps that
emit at 300\,GHz: the crucial factor here is probably the post-shock
density, which becomes too low. As existing clumps cool below
$\sim$3000\,{\sc k}, the $J=7-6$ masers disappear. We know from
observations that most of the $J=7-6$ masers were lost between early
August and mid-September 1996 at a model phase of between 0.68 and 0.77.
Model predictions (Humphreys et al.\ 1997b) suggest that for model phases $>$
0.7-0.75, the 43- and 86-GHz maser shells in $v=1$ should be similar in
size, with the 86-GHz shell being slightly the larger and less well defined.
Over a similar phase range, the peak amplifications in the synthetic maser
are in the ratio of $\sim$2:1 in favour of the 43-GHz transition. 

The original detection of the 300-GHz lines clearly belongs to an
earlier pulsational cycle than the rest of the data. These data were
taken nearly four months before the optical minimum of 1995 October 2
and should therefore correspond roughly to the 1996 June or July
observations. The fact that the 300-GHz lines were much brighter than
features detected one period later is further evidence for the
variability of SiO emission between phases, already well established
at 43 and 86\,GHz (Nyman \& Oloffson 1986; Martinez et
al.\ 1988). Evidently the earlier pulsational cycle was also better
suited to 86-GHz emission, as seen in the relative intensities of the
peaks near the beginning of 1996 October and the earliest SEST data
(which is much brighter, although almost certainly past the peak of
the previous pulsation). It is not clear from this work why any one
cycle should favour SiO maser emission, at any frequency, over any
other cycle.

\section{Concluding Remarks}

In this paper we report monitoring observations of SiO maser emission
from the $J=2-1$ and $J=7-6$ transitions and compare them with
interferometry data at $J=1-0$ and with predictions from hydrodynamic
models. We conclude that the `clump' model of maser emission is
consistent with the observations and that a plausible scenario can be
constructed for the evolution of the maser emission, both high and low
frequency, from the envelope of R~Aqr.

In the case of low-frequency masers, including both the $J=1-0$ and
$J=2-1$ data discussed in this work, the passage of a shock-wave
through the envelope both creates and destroys maser clumps. Clumps
upstream of the shock are destroyed, whilst new features appear
downstream of the shock, after a short time delay required for
cooling, provided that post-shock conditions are suitable for the
developing instability clumps to support masers.

When the shock is close to the star, the balance is in favour of
destruction, but an adequate supply of clumps remains upstream to
maintain the low-frequency spectra: it is important that these masers
can still emit near the inter-shock background temperature of the
envelope, in the range 1500---2000\,{\sc k}, whilst the 300-GHz masers
cannot.  This phase of net destruction is associated with the latter
stages of envelope contraction and the early stages of expansion.

Later, the production of new clumps behind the shock far outweighs the
loss of the few remaining upstream clumps and the low-frequency
emission passes through a maximum near a model phase of 0.85 (optical
phase $\sim$0.4).

By contrast, the high frequency ($J=7-6$) maser clumps are present
only fairly close behind the outward propagating shock: the upstream
gas is too cool to support these masers. Further cooling means that
replenishment of high-frequency emitting clumps must be sustained by
the shock and that the radial range over which the $J=7-6$ clumps can
exist is restricted: their numbers are therefore low and as the shock
weakens, the spectrum is rapidly depleted. The time range suitable for
producing $J=7-6$ emitting clumps lies roughly between model phases
0.25 and 0.7.

Most maser clumps emit in several transitions simultaneously
(Humphreys et al.\ 1996) but conditions suitable for high-frequency
emission are likely to be found only near the beginning of the
lifetime of a clump. Clumps which can emit at even higher frequency
transitions ($J=8-7$ and above) are likely to be rarer still.

\subsection*{ACKNOWLEDGMENTS}

We thank the director and staff of the James Clerk Maxwell Telescope
which is operated by the Royal Observatories on behalf of the Particle
Physics and Astronomy Research Council (PPARC) of the United Kingdom,
the Netherlands Organisation for Scientific Research and the National
Research Council of Canada. We would also like to thank Lars Nyman for
supplying the SEST data presented here. MDG acknowledges the award of
a University Research Fellowship by the Royal Society; RJI and JAY
acknowledge awards from PPARC. In this research, we have used, and acknowledge
with thanks, data from the AAVSO International Database, based on observations
submitted to the AAVSO by variable star observers worldwide.

\end{document}